\documentstyle[multicol,pre,aps,epsf,amsmath]{revtex}

\begin{document}

\draft

\title{Finite element approach for simulating quantum electron dynamics in a magnetic field}

\author{Naoki Watanabe, Masaru Tsukada}

\address{Department of Physics,%
  Graduate School of Science, University of Tokyo
  7-3-1 Hongo, 113-0033 Bunkyo-ku, Tokyo, Japan}

\date{Published from Journal of Physical Society of Japan, {\bf 69}, No.9, 2962, (2000).}

\maketitle

\begin{abstract}
A fast and stable numerical method is formulated to compute the time evolution of
a wave function in a magnetic field by solving
the time-dependent Schr{\"o}dinger equation.
This computational method is based on 
the finite element method in real space to
improved accuracy without any increase of computational cost.
This method is also based on Suzuki's exponential product theory
to afford an efficient way to manage the TD-Schr{\"o}dinger
equation with a vector potential.
Applying this method to some simple electron dynamics,
we have confirmed its efficiency and accuracy.
\end{abstract}


\begin{multicols}{2}
\narrowtext

\section{Introduction}

Conventionally, wave functions have been represented as a linear combination of
plane waves or atomic orbitals in the calculations of
the electronic states or their time evolution.
However, these representations
entail high computational cost to calculate the matrix elements for these bases.
The plane wave bases set is not suitable for localized orbitals, and the atomic
orbital bases set is not suitable for spreading waves. 

To overcome those problems, some numerical methods adopted
real-space representation to solve the time dependent Schr{\"o}dinger equation
\cite{Varga1962,DeRaedt1994,Iitaka1994,Natori1997}.
In those methods, a wavefunction is descritized by grid points
in real space and the spatial differential operator is approximated
by the finite difference method (FDM).
With those methods, some dynamic electron phenomena were simulated successfully
\cite{DeRaedt1994PRB,Iitaka1997,Kono1997}.

In the previous work\cite{Watanabe2000}, we have formulated
a new computational method for the TD-Schr{\"o}dinger equation by using some
computational techniques such as, the FDM, Suzuki's exponential product theory
\cite{Suzuki1990,Suzuki1991,Umeno1993,Suzuki1993,Suzuki1993springer,Suzuki1995},
Cayley's form\cite{Recipes} and Adhesive operator.
This method afforded high-stability and low computational cost.

In the field of
engineering, for example, numerical analysis of fluid dynamics or of
strength of macroscopic constructions,
the finite element method (FEM) has been widely and traditionally used
for approximating the appropriate partial differential equations.
Recently, the FEM has been found useful for
the time-independent Schr{\"o}dinger equation of electrons
in solid or liquid materials\cite{Tsuchida1998}.

In this paper, we have utilized the FEM for solving the TD-Schr{\"o}dinger
equation as an extension of the previous work\cite{Watanabe2000}.
By using Cayley's form and the FEM,
this method affords high-accuracy without any increase of computational cost.
Moreover, we have formulated a new efficient method which manages
the time evolution of a wave function in a vector potential
or in a magnetic field.
These techniques are especially useful for simulating dynamics of electrons
in a variety of meso-scopic systems.

\section{Formulation}

In this section, we formulate a new method derived by the FEM and 
a new scheme to manage a vector potential efficiently.
Throughout this paper, we often use the atomic unit $\hbar=1,\,m=1,\,e=1$.

\subsection{FEM for the TD-Schr{\"o}dinger equation}

First, we utilize the FEM for
the time evolution of a wave function in a one-dimensional closed system
described by the following TD-Schr{\"o}dinger equation:
\begin{equation}
  {\rm i} \hbar \frac{\partial\psi(x,t)}{\partial t}
= 
  -\frac{\hbar^2}{2m} \frac{\partial^2}{\partial x^2}\, \psi(x,t)\ .
  \label{2-1-1}
\end{equation}

\begin{figure}[H]
  \begin{center}
    \epsfxsize40mm\mbox{\epsfbox{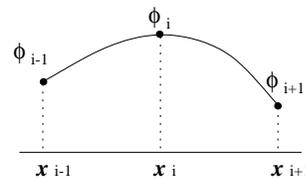}}
  \end{center}
  \caption{
    The FEM starts by smoothing the wavefunction around a grid point.
    The wavefunction is supplemented by a quadratic equation.
  }
  \label{fig2-1-1}
\end{figure}

The FEM starts by smoothing the wavefunction around a grid point.
We smoothed $\psi(x)$ around a grid point $x_i$ by eq.~(\ref{2-1-2}),
as illustrated in Fig.~\ref{fig2-1-1}:
\begin{equation}
  \psi(x,t)
  =
  \psi_i(t) + \psi_i^\prime(t) (x-x_i)
  + \frac{1}{2} \psi_i^{\prime\prime}(t) (x-x_i)^2\ ,
  \label{2-1-2}
\end{equation}
where
\begin{equation}
\begin{split}
  \psi_i(t)
  &\equiv
  \psi(x_i,t)\ , \\
  \psi_i^\prime(t)
  &\equiv
  \frac{\psi_{i+1}(t)-\psi_{i-1}(t)}{2\Delta{x}}\ , \\
  \psi_i^{\prime\prime}(t)
  &\equiv
  \frac{\psi_{i+1}(t)-2\psi_{i}(t)+\psi_{i-1}(t)}{\Delta{x}^2}\ .
\end{split}
  \label{2-1-3}
\end{equation}

By substituting eq.~(\ref{2-1-3}) for eq.~(\ref{2-1-2}),
$\psi(x,t)$ is expressed as
\begin{equation}
  \psi(x,t)
=
    u_{{\rm a}i}(x)\, \psi_{i-1}(t)
  + u_{{\rm o}i}(x)\, \psi_{i}(t)
  + u_{{\rm b}i}(x)\, \psi_{i+1}(t)\ ,
  \label{2-1-4}
\end{equation}
where $u_{{\rm a}i}(x),\, u_{{\rm o}i}(x)$ and $u_{{\rm b}i}(x)$
are the base functions defined below:
\begin{equation}
\begin{split}
  u_{{\rm a}i}(x)
  &=
  \frac{(\!x\!-\!x_i\!)^2}{2\Delta{x}^2}
- \frac{(\!x\!-\!x_i\!)}{2\Delta{x}}\ , \\
  u_{{\rm o}i}(x)
  &=
  1 - \frac{(\!x\!-\!x_i\!)^2}{\Delta{x}^2}\ , \\
  u_{{\rm b}i}(x)
  &=
  \frac{(\!x\!-\!x_i\!)^2}{2\Delta{x}^2}
+ \frac{(\!x\!-\!x_i\!)}{2\Delta{x}}\ .
\end{split}
  \label{2-1-5}
\end{equation}

Substituting eq.~(\ref{2-1-4}) for eq.~(\ref{2-1-1}) and
multiplying both side of the equation by the base function $u_{{\rm o}i}(x)$
and integrating by $x$ in the range $[x_{i-1},\,x_{i+1}]$ as
\begin{align}
&  {\rm i} \hbar \int_{x_{i-1}}^{x_{i+1}} \!\!\!\! {\rm d}x\,
    u_{{\rm o}i}(x) \Bigl[
      u_{{\rm o}i}(x) \,\dot{\psi}_{i  }(t) \notag\\
&\hskip10mm    + u_{{\rm a}i}(x) \,\dot{\psi}_{i+1}(t)
    + u_{{\rm b}i}(x) \,\dot{\psi}_{i-1}(t)
  \Bigr] \notag\\
=
&  -\frac{\hbar^2}{2m} \int_{x_{i-1}}^{x_{i+1}} \!\!\!\! {\rm d}x\,
    u_{{\rm o}i}(x) \Bigl[
      \partial_x^2 u_{{\rm o}i}(x) \,\psi_{i  }(t) \notag\\
&\hskip10mm
    + \partial_x^2 u_{{\rm a}i}(x) \,\psi_{i+1}(t)
    + \partial_x^2 u_{{\rm b}i}(x) \,\psi_{i-1}(t)
  \Bigr]\ ,
  \label{2-1-6}
\end{align}
the following formula is obtained after some algebra:
\begin{align}
 & {\rm i} \hbar \frac{1}{10} \bigl[
    \dot{\psi}_{i-1}(t) + 8 \dot{\psi}_{i}(t) + \dot{\psi}_{i+1}(t)
  \bigr] \notag\\
=
&
  -\frac{\hbar^2}{2m\Delta{x}^2} \bigl[
    \psi_{i-1}(t) - 2 \psi_{i}(t) + \psi_{i+1}(t)
  \bigr] \ .
  \label{2-1-7}
\end{align}

To simplify the expression, it is useful to define
a vector and two matrices as below:
\begin{equation}
  \psi(t) \equiv (\psi_0(t),\ldots,\psi_{N-1}(t))^{\rm T}\ ,
  \label{2-1-8}
\end{equation}
\begin{equation}
\begin{split}
  {\bf S}
&\equiv
  \frac{1}{10}
  \left[\begin{array}{cccc}
    8 & 1 & 0 & 0 \\
    1 & \ddots & \ddots & 0 \\
    0 & \ddots & \ddots & 1 \\
    0 & 0 & 1 & 8
  \end{array}\right], \\
  {\bf D}
&\equiv
  \left[\begin{array}{cccc}
    -2 & 1 & 0 & 0 \\
    1 & \ddots & \ddots & 0 \\
    0 & \ddots & \ddots & 1 \\
    0 & 0 & 1 & -2
  \end{array}\right].
\end{split}
  \label{2-1-9}
\end{equation}
Clearly ${\bf S}$ and ${\bf D}$ satisfy the following equation:
\begin{equation}
  {\bf S}
=
  {\bf I} + \frac{1}{10} {\bf D}\ .
  \label{2-1-10}
\end{equation}
Using these notations, eq.~(\ref{2-1-7}) is expressed simply as
\begin{equation}
  {\rm i} \hbar {\bf S} \frac{\partial\psi(t)}{\partial t}
=
  - \frac{\hbar^2}{2m\Delta{x}^2} {\bf D}\ \psi(t)\ .
  \label{2-1-11}
\end{equation}
Equation~(\ref{2-1-11}) is the finite element equation for this case.

It has been thought that the existence of the matrix ${\bf S}$ is
troublesome since the inverse of this matrix is required to
obtain the time derivative of the wave function, namely,
\begin{equation}
  \frac{\partial\psi(t)}{\partial t}
  =
  \frac{{\rm i}\hbar}{2m\Delta{x}^2} {\bf S}^{-1} {\bf D}\ \psi(t)\ .
  \label{2-1-12}
\end{equation}

However, we have found that this differential equation is
easily solved by using an approximation called Cayley's form.
The formal solution of eq.~(\ref{2-1-12}) is given by,
\begin{equation}
  \psi(t+\Delta{t})
=
  \exp{\Bigl[
    \frac{{\rm i}\hbar}{2m} \frac{\Delta{t}}{\Delta{x}^2} {\bf S}^{-1}{\bf D}
  \Bigr]} \  \psi(t)\ .
  \label{2-1-13}
\end{equation}
The exponential operator is approximated by Cayley's form:
\begin{equation}
  \psi(t+\Delta{t})
  \simeq
  \frac{\displaystyle {\bf I} + \frac{{\rm i}\hbar}{4m}
            \frac{\Delta{t}}{\Delta{x}^2} {\bf S}^{-1}{\bf D}}
      {\displaystyle {\bf I} - \frac{{\rm i}\hbar}{4m}
            \frac{\Delta{t}}{\Delta{x}^2} {\bf S}^{-1}{\bf D}}\ 
       \psi(t)\ .
  \label{2-1-14}
\end{equation}
Multiplying both the numerator and the denominator of the
righthand side by the matrix ${\bf S}$ and using the relation (\ref{2-1-6}),
the required formula is obtained:
\begin{equation}
  \psi(t+\Delta{t})
=
  \frac{\displaystyle {\bf I} + \frac{{\rm i}\hbar}{4m_{\rm eff}}
            \frac{\Delta{t}}{\Delta{x}^2} {\bf D}}
       {\displaystyle {\bf I} - \frac{{\rm i}\hbar}{4m_{\rm eff}^\ast}
            \frac{\Delta{t}}{\Delta{x}^2} {\bf D}}\ 
       \psi(t)\ ,
  \label{2-1-15}
\end{equation}
where $m_{\rm eff}$ is an ``effective mass'' of an electron
defined as
\begin{equation}
  \frac{\hbar}{m_{\rm eff}}
\equiv
  \frac{\hbar}{m} - {\rm i}\frac{2}{5} \frac{\Delta{x}^2}{\Delta{t}}\ .
  \label{2-1-16}
\end{equation}

In this way, the solution of the partial differential equation,
eq.~(\ref{2-1-1}) is computed
by eq.~(\ref{2-1-15}) with the concept of the FEM.
It is quite a remarkable result that formula eq.~(\ref{2-1-15}) 
is almost the same as the formula derived by the FDM\cite{Watanabe2000}.
In this time evolution, the norm of the wave function is exactly
conserved since the time evolution operator appearing in eq.~(\ref{2-1-1})
is strictly unitary.
Moreover, accuracy is dramatically improved without any increase in
the computational cost, as demonstrated in the next section.


It is easy to extend this idea for two-dimensional systems,
since the time evolution operator in a two-dimensional system
is decomposed into a product of the time evolution operators in one-dimensional
systems\cite{Watanabe2000}.
The approximated solution utilizing the FEM is given by
\begin{equation}
  \psi({\bf r},t+\Delta{t})
  =
  \frac{\displaystyle {\bf I} + \frac{{\rm i}\hbar}{4m_{\rm eff}}
            \frac{\Delta{t}}{\Delta{x}^2} {\bf D}_x}
       {\displaystyle {\bf I} - \frac{{\rm i}\hbar}{4m_{\rm eff}^\ast}
            \frac{\Delta{t}}{\Delta{x}^2} {\bf D}_x} \cdot
  \frac{\displaystyle {\bf I} + \frac{{\rm i}\hbar}{4m_{\rm eff}}
            \frac{\Delta{t}}{\Delta{y}^2} {\bf D}_y}
       {\displaystyle {\bf I} - \frac{{\rm i}\hbar}{4m_{\rm eff}^\ast}
            \frac{\Delta{t}}{\Delta{y}^2} {\bf D}_y}
\,
  \psi({\bf r},t)\ ,
\label{2-1-22}
\end{equation}
where ${\bf D}_x$ and ${\bf D}_y$ are
the finite difference matrices along the $x$ and $y$ axes respectively,
and their appearances are the same as ${\bf D}$ defined in eq.~(\ref{2-1-9}).

\subsection{Evolution in a magnetic field}

Though there are many interesting phenomena in a magnetic field,
there has been no efficient methods that numerically manage the dynamics in
a magnetic field as far as we know.
We have improved our method to afford an efficient way to
solve the TD-Schr{\"o}dinger equation with a vector potential given as below
\begin{equation}
  {\rm i} \hbar \frac{\partial\psi({\bf r},t)}{\partial t}
=
  - \frac{\hbar^2}{2m}
    \Bigl( \nabla - \frac{{\rm i}e}{\hbar} {\bf A} \Bigr)^2
  \psi({\bf r},t)\ .
  \label{2-2-1}
\end{equation}

In this subsection, we present the method for only
the case of a two-dimensional system lying on the $xy$ plane
subjected to a uniform external magnetic field along the $z$ axis.
We do not mention the case of a non-uniform magnetic field specifically,
but the extension of the method is straightforward.
We adopt the following vector potential ${\bf A}$
for this magnetic field:
\begin{equation}
  {\bf A}
=
  ( -By,\, 0,\, 0 )^{\rm T}\ .
  \label{2-2-2}
\end{equation}
The TD-Schr{\"o}dinger equation of this system is given by
\begin{equation}
  {\rm i} \hbar \frac{\partial\psi({\bf r},t)}{\partial t}
=
  \Bigr[
  - \frac{\hbar^2}{2m}
    \Bigl( \partial_x - \frac{{\rm i}e}{\hbar} By \Bigr)^2
  - \frac{\hbar^2}{2m} \partial_y^2
  \Bigr]
  \psi({\bf r},t)\ .
  \label{2-2-3}
\end{equation}
The strict, analytical solution is also given by an exponential operator:
\begin{equation}
  \psi({\bf r}, t+\Delta{t})
=
  \exp \Bigl[
    \frac{{\rm i}\hbar}{2m}\Delta{t}
    \Bigl(
      \partial_x
    - \frac{{\rm i}e}{\hbar} By
    \Bigr)^2
+ \frac{{\rm i}\hbar}{2m}\Delta{t}
    \partial_y^2
  \Bigr]\
  \psi({\bf r}, t)\ .
  \label{2-2-4}
\end{equation}

Note the following identity:
\begin{gather}
  \exp{\left[
  \Delta{t}\frac{{\rm i}\hbar}{2m}
  \Bigl(
    \partial_x - \frac{{\rm i}e}{\hbar} By
  \Bigr)^2
  \right]}
= 
  \exp{\Bigl(+\frac{{\rm i}e}{\hbar} Bxy \Bigr)} \notag\\
\hskip10mm
  \times
  \exp{\left[\Delta{t}\frac{{\rm i}\hbar}{2m}
      \partial_x^2
  \right]}
  \exp{\Bigl(-\frac{{\rm i}e}{\hbar} Bxy \Bigr)}\ .
  \label{2-2-5}
\end{gather}

Equation (\ref{2-2-4}) is approximated
by the following second-order exponential product:
\begin{multline}
  \psi({\bf r}, t+\Delta{t})
=
  \exp{\left[\frac{\Delta{t}}{2}\frac{{\rm i}\hbar}{2m}
      \partial_y^2
  \right]}
  \exp{\Bigl(+\frac{{\rm i}e}{\hbar} Bxy \Bigr)} \\
\times
  \exp{\left[\Delta{t}\frac{{\rm i}\hbar}{2m}  
      \partial_x^2
  \right]}
  \exp{\Bigl(-\frac{{\rm i}e}{\hbar} Bxy \Bigr)} \\
\times
  \exp{\left[\frac{\Delta{t}}{2}\frac{{\rm i}\hbar}{2m}
      \partial_y^2
  \right]}
  \psi({\bf r}, t)
  + O(\Delta{t}^3)\ .
  \label{2-2-6}
\end{multline}

Moreover, we have found that the hybrid decomposition\cite{Suzuki1995}
is rather easy in this case. Note the following identity:
\begin{equation}
  \left[ \partial_y^2,
    \left[ ( \partial_x - {\rm i}a y )^2,
         \partial_y^2 \right]
  \right]
  =
  - 8 a^2 \partial_y^2\ .
  \label{2-2-7}
\end{equation}

Then, equation (\ref{2-2-4}) is approximated
by the following fourth-order hybrid exponential product:
\begin{multline}
  \psi({\bf r}, t+\Delta{t})
=
  \exp{\Bigl[ \Delta{t}\frac{{\rm i}\hbar}{2m} \Bigl(
    \frac{1}{6} - \frac{e^2B^2\Delta{t}^2}{72 m^2c^2} 
  \Bigr) \partial_y^2 \Bigr]} \\
\times
  \exp{\Bigl(+\frac{{\rm i}e}{\hbar} Bxy \Bigr)}
  \exp{\Bigl[ \frac{\Delta{t}}{2} \frac{{\rm i}\hbar}{2m} \partial_x^2 \Bigr]}
  \exp{\Bigl(-\frac{{\rm i}e}{\hbar} Bxy \Bigr)} \\
\times
  \exp{\Bigl[ \frac{2\Delta{t}}{3} \frac{{\rm i}\hbar}{2m} \partial_y^2 \Bigr]}\\
\times
  \exp{\Bigl(+\frac{{\rm i}e}{\hbar} Bxy \Bigr)}
  \exp{\Bigl[ \frac{\Delta{t}}{2} \frac{{\rm i}\hbar}{2m} \partial_x^2 \Bigr]}
  \exp{\Bigl(-\frac{{\rm i}e}{\hbar} Bxy \Bigr)} \\
\times
  \exp{\Bigl[ \Delta{t}\frac{{\rm i}\hbar}{2m} \Bigl(
    \frac{1}{6} - \frac{e^2B^2\Delta{t}^2}{72 m^2c^2} 
  \Bigr) \partial_y^2 \Bigr]}
  \psi({\bf r}, t)
  + O(\Delta{t}^5)\ .
  \label{2-2-8}
\end{multline}

\begin{figure}[H]
  \begin{center}
    \epsfxsize=80mm\mbox{\epsfbox{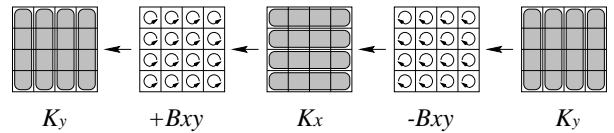}}
  \end{center}
  \caption{
    The procedure for a two-dimensional system subjected to a magnetic field.
    Here $Bxy$ means the operation of the exponential of the magnetic field.
    In this way, the phase of the wavefunction is turned forward before
    the operation of Cayley's form along the x-axis
    and is turned backward after Cayley's form.
  }
  \label{fig2-2-1}
\end{figure}

The exponential of the magnetic field just changes the phase of the wave
function, so it is very easy to compute.
Therefore, this method is adaptable to systems subjected to a magnetic field.
The outline of the procedure for a two-dimensional 
system subjected to a magnetic field is schematically described by Fig.~\ref{fig2-2-1}.

\section{Applications}

\subsection{Comparison between FDM and FEM}

\begin{figure}[H]
  \begin{center}
    \epsfxsize=80mm\mbox{\epsfbox{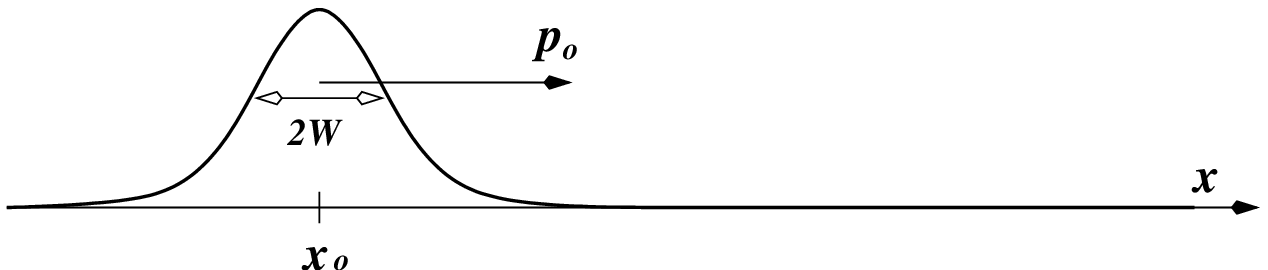}}%
  \end{center}
  \caption{
    The model system for comparison with the conventional methods.
    $256$ computational grid points are allocated in the physical length
    $8.0 \text{a.u.}$
    A Gaussian wave packet is placed in the system, whose initial
    average location $x_o$ and momentum $p_o$ are set as
    $x_o=2.0\text{a.u.}$ and $p_o=12.0\text{a.u.}$, respectively.
  }
  \label{fig3-1-1}
\end{figure}

In this subsection, we briefly compare Cayley's form
and other conventional methods
by simply simulating a Gaussian wave packet moving in a one-dimensional
free system as illustrated in Fig.~\ref{fig3-1-1}.

The TD-Schr{\"o}dinger equation of this system is simply given by
\begin{equation}
  {\rm i}\frac{\partial \psi(x,t)}{\partial t}
=
  - \frac{\partial_x^2}{2}\ \psi(x,t)\ .
  \label{3-1-1}
\end{equation}
The wavefunction at the initial state is set as a Gaussian:
\begin{equation}
  \psi(x,t=0)
=
  \frac{1}{\sqrt[4]{2\pi W^2}}
  \exp{\Bigl[
     -\frac{|x-x_o|^2}{4W^2} + {\rm i} p_o x
  \Bigr]}\ ,
  \label{3-1-2}
\end{equation}
where $W=0.25\text{a.u.}\ x_o=2.0\text{a.u.}\ p_o=12.0\text{a.u.}$
The evolution of this Gaussian is analytically derived as
\begin{gather}
  \psi(x,t)
=
  \frac{1}{\sqrt[4]{2\pi W^2 + (\pi/2)(t/W)^2}} \notag\\
\hskip10mm
  \times
  \exp{\Bigl[
    - \frac{(x-x_o-p_o t)^2}{4W^2+(t/W)^2}
    + {\rm i} p_o x
  \Bigr]}\ .
  \label{3-1-3}
\end{gather}
Therefore, the average location of the Gaussian $\langle x(t) \rangle$
is derived as if it is a classical particle:
\begin{equation}
  \langle x(t) \rangle 
=
  \langle x(t=0) \rangle + p_o t\ .
  \label{3-1-4}
\end{equation}
This characteristic is useful to check the accuracy of the simulation.

Cayley's form with the FDM is given by
\begin{equation}
  \psi(t+\Delta{t})
=
  \frac{1 + {\rm i}\Delta{t}/4\ \partial_{x}^2}
       {1 - {\rm i}\Delta{t}/4\ \partial_{x}^2}\,
  \psi(t)\ ,
  \label{3-1-5}
\end{equation}
where $\partial_x^2$ is approximated by a finite difference matrix as
\begin{equation}
  \partial_{x}^2
  \simeq
  \frac{1}{\Delta{x}^2}
  \left[\begin{array}{cccccc}
   -2 & 1 & 0 & 0 & 0 & 0 \\
    1 &-2 & 1 & 0 & 0 & 0 \\
    0 & 1 &-2 & 1 & 0 & 0 \\
    0 & 0 & 1 &-2 & 1 & 0 \\
    0 & 0 & 0 & 1 &-2 & 1 \\
    0 & 0 & 0 & 0 & 1 &-2
  \end{array}\right]\ .
  \label{3-1-6}
\end{equation}

Meanwhile, Cayley's form with the FEM is given by
\begin{equation}
  \psi(t+\Delta{t})
=
  \frac{m_{\rm eff}      + {\rm i}\Delta{t}/4 \partial_{x}^2}
       {m_{\rm eff}^\ast - {\rm i}\Delta{t}/4 \partial_{x}^2}\,
  \psi(t)\ ,
  \label{3-1-7}
\end{equation}
where the spatial differential operator is approximated in the ordinary way
and $m_{\rm eff}$ is the effective mass:
\begin{equation}
  \frac{1}{m_{\rm eff}}
  \equiv
  \frac{1}{m} - {\rm i}\frac{2}{5} \frac{\Delta{x}^2}{\Delta{t}}\ ,
  \label{3-1-8}
\end{equation}
$\partial_x^2$ is approximated by eq.~(\ref{3-1-6}).

\begin{figure}[H]
  \begin{center}
    \epsfxsize80mm\mbox{\epsfbox{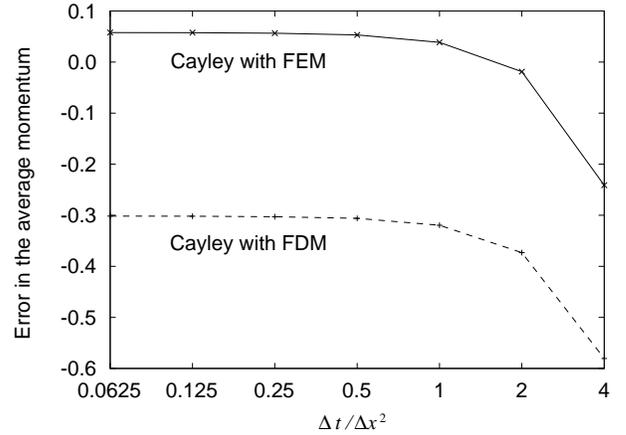}}
  \end{center}
  \caption{%
    The errors in the average momentum computed by Cayley's form
    with the FDM and Cayley's form with the FEM.
    The error of the FEM is smaller than that of the FDM.
    The spatial slice is set as $\Delta{x}=1/32\text{a.u.}$
  }
  \label{fdmfem}
\end{figure}

We have simulated the motion of the Gaussian by those methods.
Figure~\ref{fdmfem} shows the error in the average momentum.
The errors are evaluated in the following way:
\begin{gather}
  \epsilon(\Delta{t}/\Delta{x}^2)
  =
  \frac{\langle x(t=T) \rangle - x_o}{T} - p_o\ ,
  \label{3-1-9} \\
  \langle x(t) \rangle
  =
  \Delta{x} \sum_{i=0}^{N-1}
  x_i |\psi_i(t)|^2
  \quad \text{in FDM.} \\
  \langle x(t) \rangle
  =
  \frac{\Delta{x}}{30} {\rm Re} \sum_{i=0}^{N-1}
  x_i \psi_i^{\ast}
  \bigl(
   24 \psi_{i}
  + 4 \psi_{i+1} + 4 \psi_{i-1} \notag\\
  -   \psi_{i+2} - \psi_{i-2}
  \bigr)
  \quad \text{in FEM.}
\end{gather}

It is found that the accuracy is dramatically improved by using the FEM.
It is remarkable that in spite of the improvement of accuracy,
the computational cost does not increase at all.

\subsection{Cyclotron motion}

\begin{figure}[H]
  \begin{center}
    \epsfxsize=36mm\mbox{\epsfbox{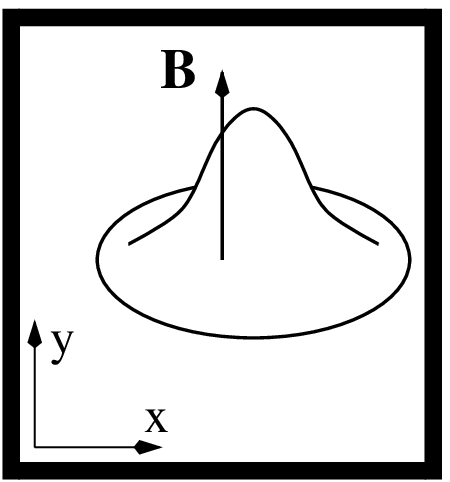}}%
  \end{center}
  \caption{
    The model system for the cyclotron motion.
    This system is subjected to a static magnetic force perpendicularly.
    and it is surrounded by infinitely high potentials.
    $64\times 64$ computational grid points are allocated in the physical length
    $8\text{a.u.}\times 8\text{a.u.}$
    The strength of the static magnetic force $B$ is set as $2\text{a.u.}$
    A Gaussian is placed as the initial state of the wavefunction,
    whose average location and momentum are set as $(6\text{a.u.},4\text{a.u.})$
    and $(0\text{a.u.}, 4\text{a.u.})$, respectively.
    The time slice is set as $\Delta{t}=1/64\text{a.u.}$
  }
  \label{fig3-2-1}
\end{figure}

We demonstrate the cyclotron motion in the framework of quantum mechanics.
We have simulated the motion of a Gaussian wave packet in
a uniform magnetic force as illustrated in Fig.~\ref{fig3-2-1}.

The initial wavefunction $\psi({\bf r},t=0)$ is set as the following 
Gaussian:
\begin{equation}
  \psi({\bf r},t=0)
  =
  \frac{1}{\sqrt{2\pi W^2}}
  \exp{\Bigl[ -\frac{|{\bf r}-{\bf r_o}|^2}{4W^2} \Bigr]}
  \exp{\Bigl[ \frac{{\rm i}eB}{\hbar} (x-L/2) y \Bigr]}\  ,
  \label{3-2-1}
\end{equation}
where ${\bf r_o}$ is set as $x_o=6\text{a.u.}, y_o=4\text{a.u.}$
and $W$ is set as $0.5\text{a.u.}$

The initial density $\rho({\bf r},t=0)$ and the initial
current density ${\bf j}({\bf r},t=0)$ derived from this wave
function are as follows:
\begin{gather}
  \rho({\bf r},t=0)
  =
  \frac{1}{2\pi W^2}
  \exp{\Bigl[-\frac{ |{\bf r}-{\bf r_o}|^2}{2W^2}\Bigr]}\  ,
  \label{3-2-2}\\
  {\bf j}({\bf r},t=0)
  =
  \frac{e^2B\rho({\bf r})}{mc} ( 0,\, x-L/2,\, 0 )^{\rm T}\ .
  \label{3-2-3}
\end{gather}

We adopt a gauge of the vector potential ${\bf A}$ as
\begin{equation}
  {\bf A}
=
  ( -By,\, 0,\, 0 )^{\rm T}\ .
  \label{3-2-4}
\end{equation}

In classical mechanics, the average momentum of this
Gaussian at the initial state is evaluated as
\begin{equation}
  p_o
=
  \frac{m}{e} | \langle {\bf j} \rangle |
=
  \frac{eB}{c} |x_o-L/2| \  .
  \label{3-2-5}
\end{equation}
This means the classical cyclotron radius is $|x_o-L/2|$.

\begin{figure}[H]
  \begin{center}\parindent0mm
    \begin{tabular}{cccc}
      \epsfxsize=20mm\mbox{\epsfbox{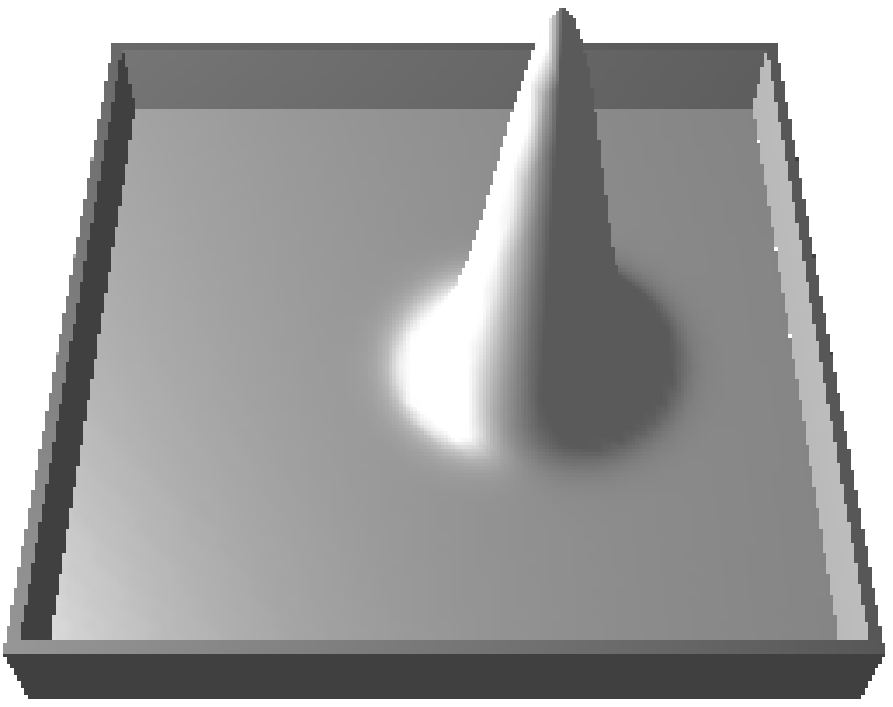}} &
      \epsfxsize=20mm\mbox{\epsfbox{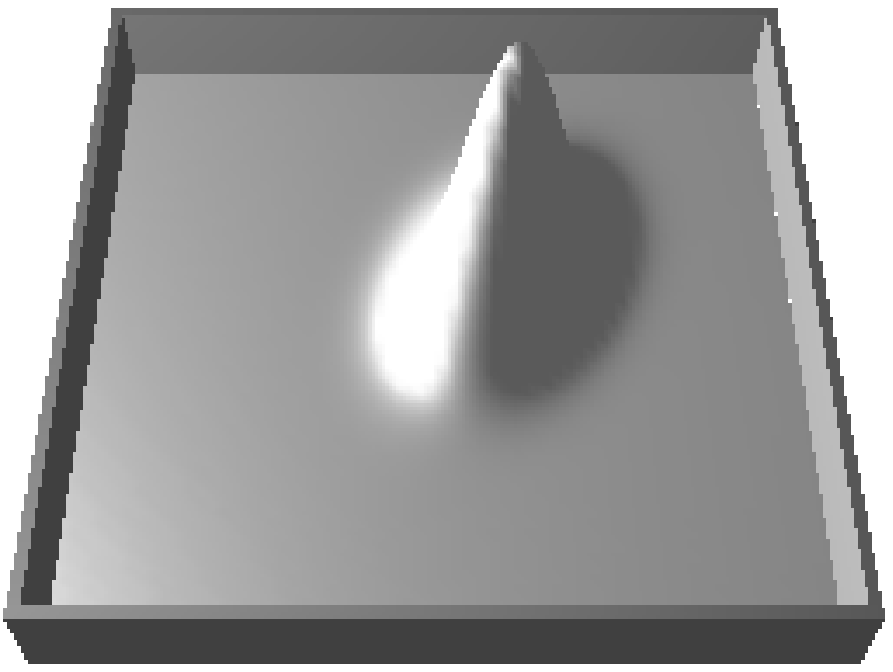}} &
      \epsfxsize=20mm\mbox{\epsfbox{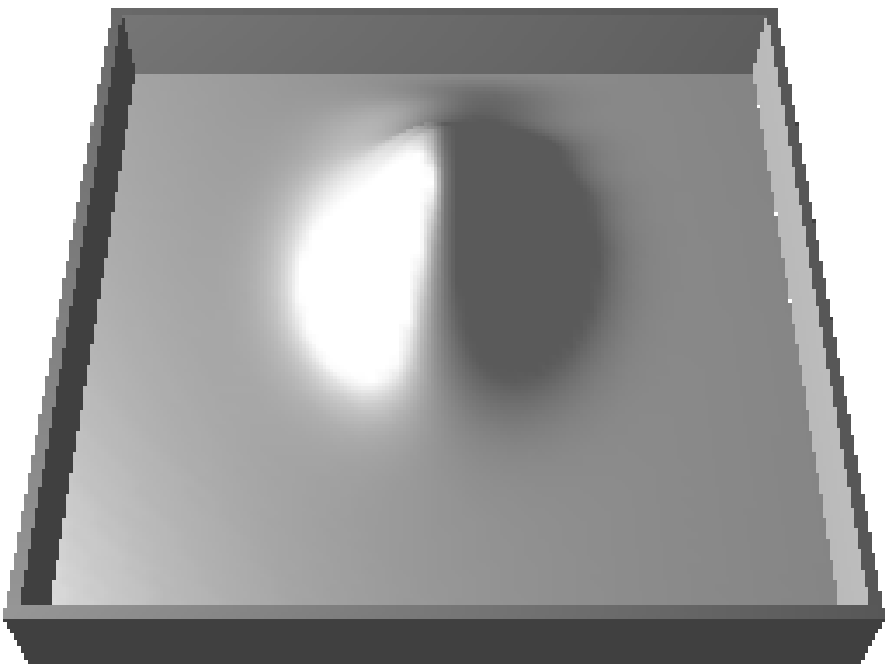}} &
      \epsfxsize=20mm\mbox{\epsfbox{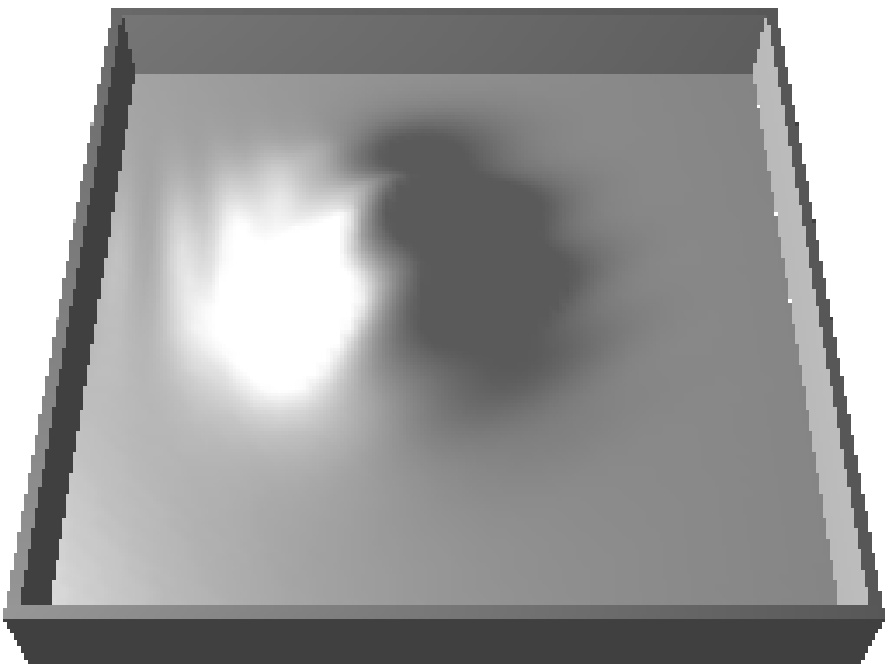}} \\
      \epsfxsize=20mm\mbox{\epsfbox{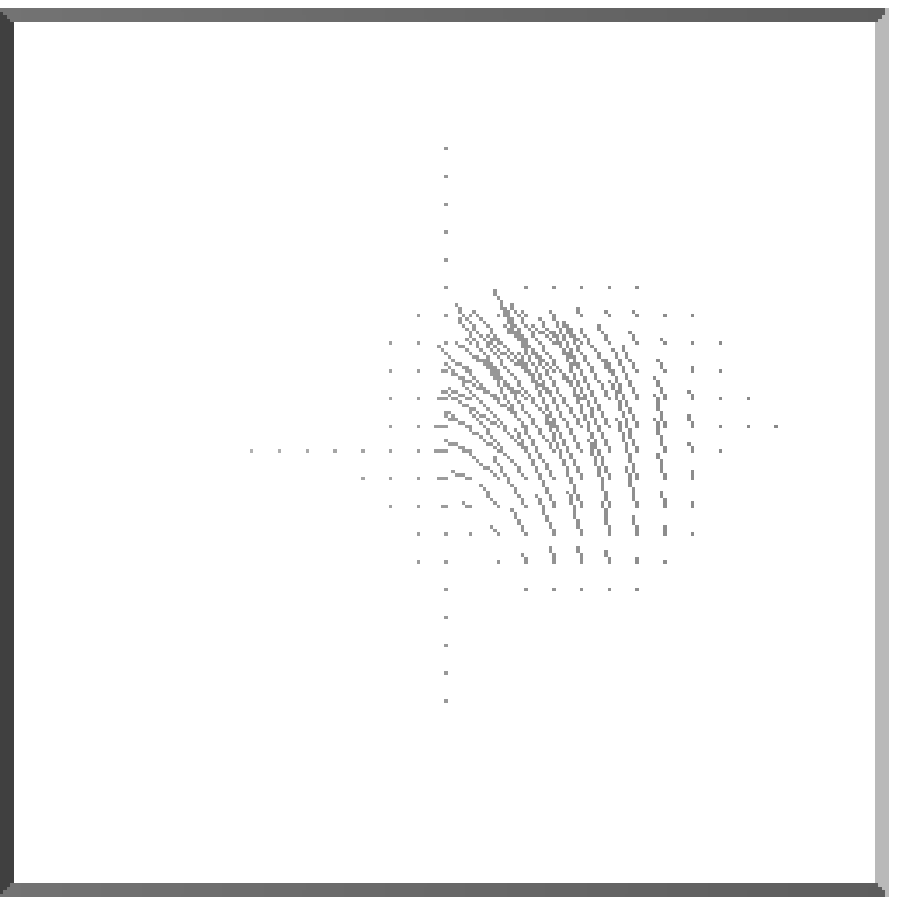}} &
      \epsfxsize=20mm\mbox{\epsfbox{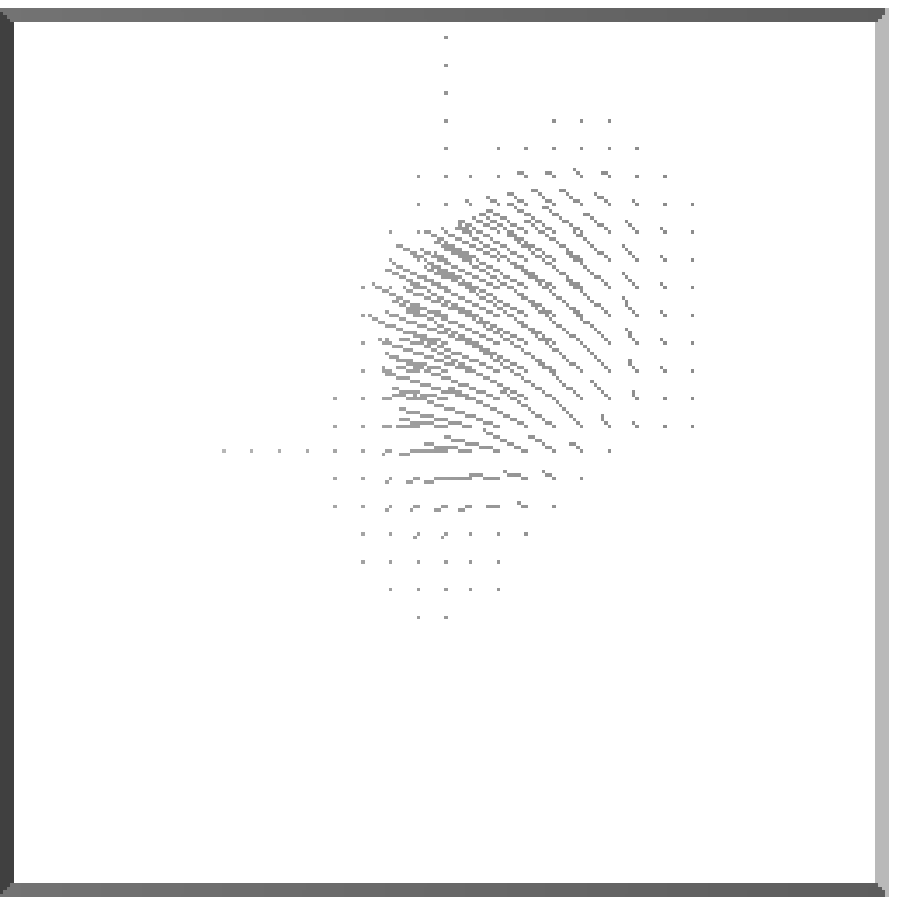}} &
      \epsfxsize=20mm\mbox{\epsfbox{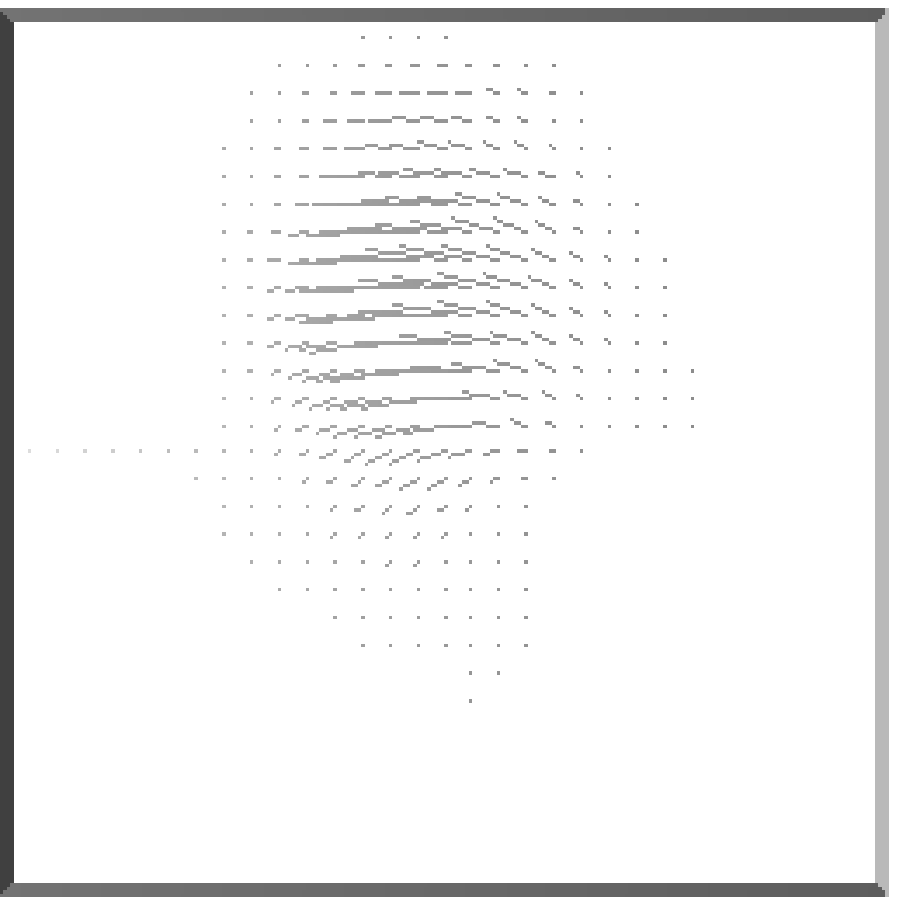}} &
      \epsfxsize=20mm\mbox{\epsfbox{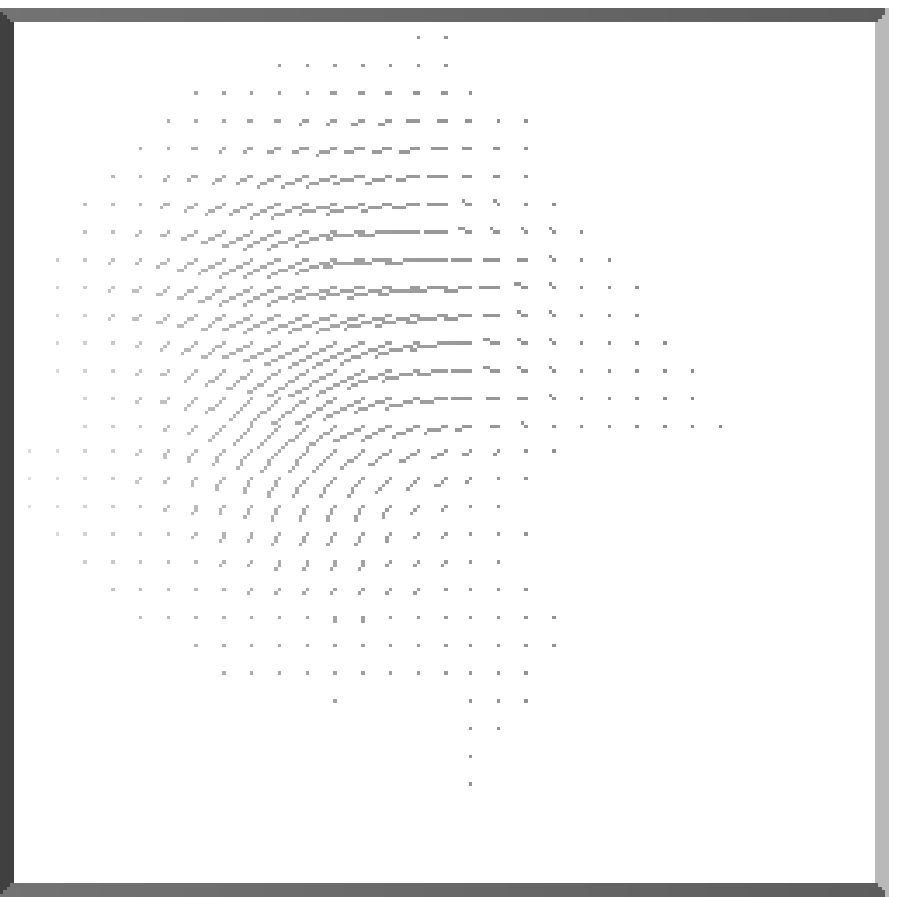}} \\
      t=0 & t=3/8 & t=6/8 & t=9/8 \\\\
      \epsfxsize=20mm\mbox{\epsfbox{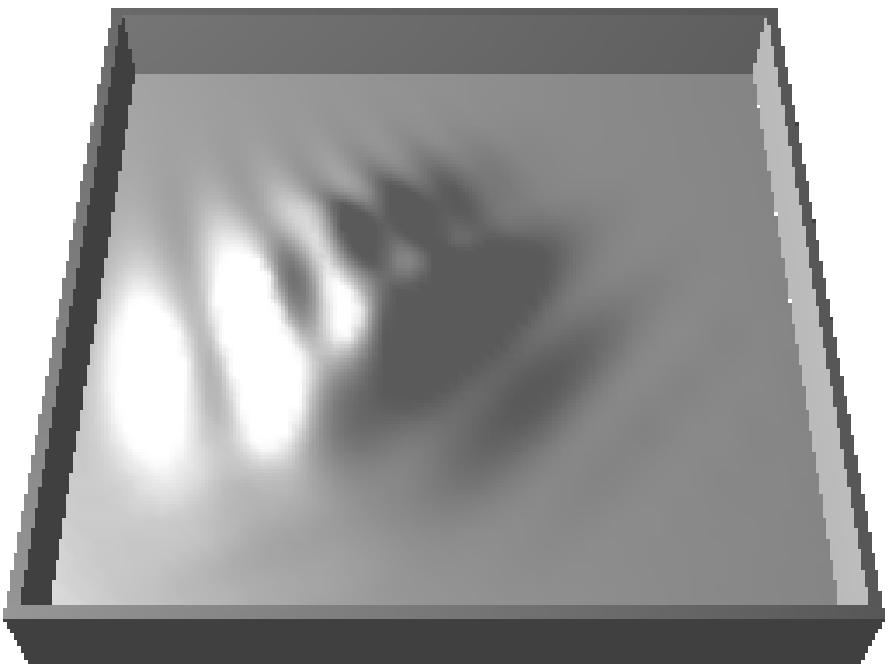}} &
      \epsfxsize=20mm\mbox{\epsfbox{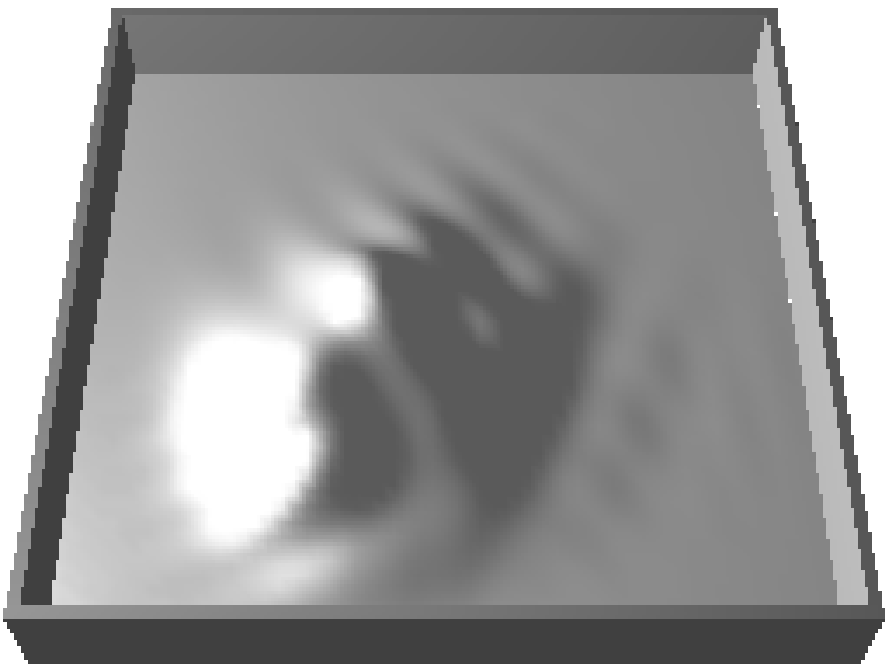}} &
      \epsfxsize=20mm\mbox{\epsfbox{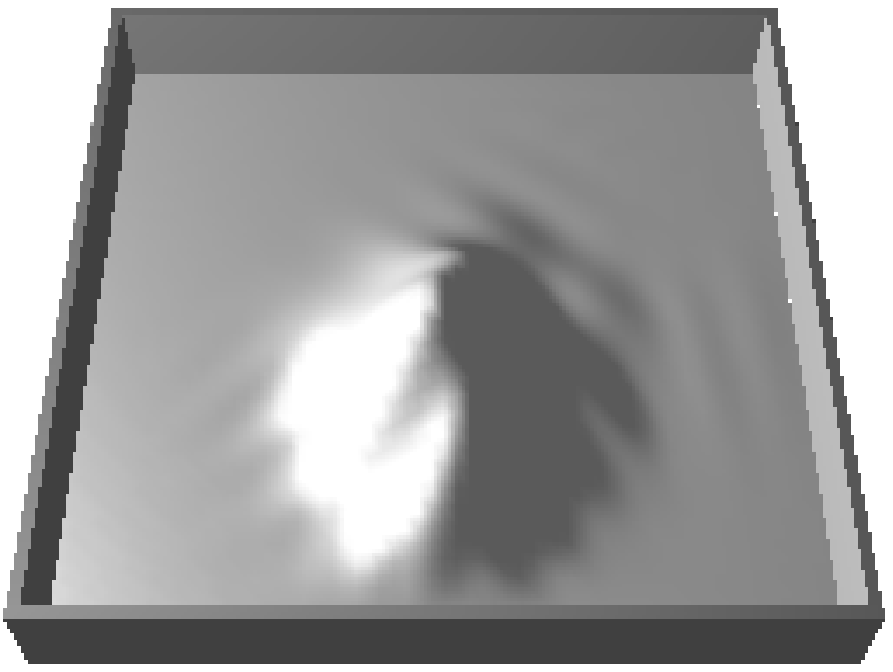}} &
      \epsfxsize=20mm\mbox{\epsfbox{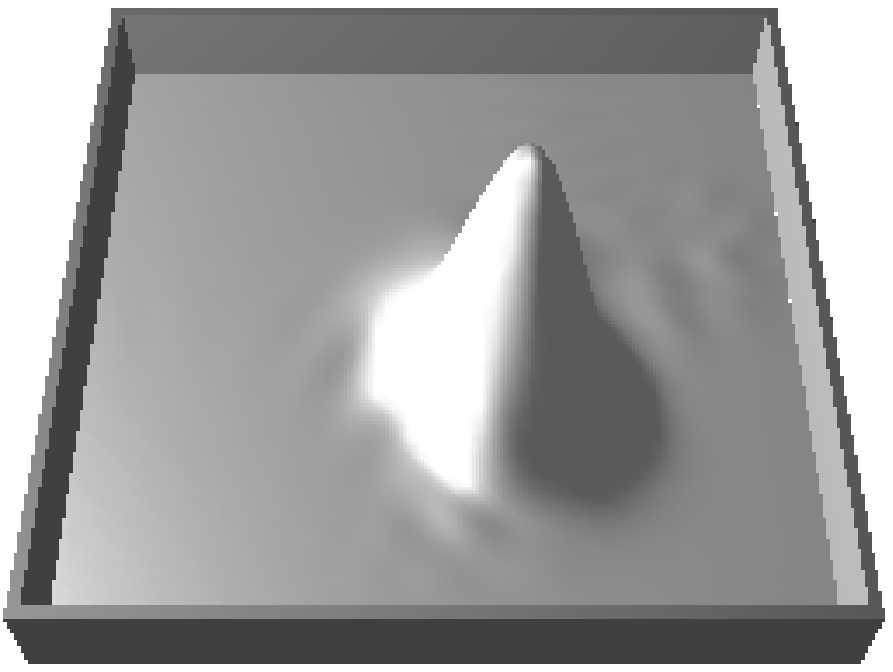}} \\
      \epsfxsize=20mm\mbox{\epsfbox{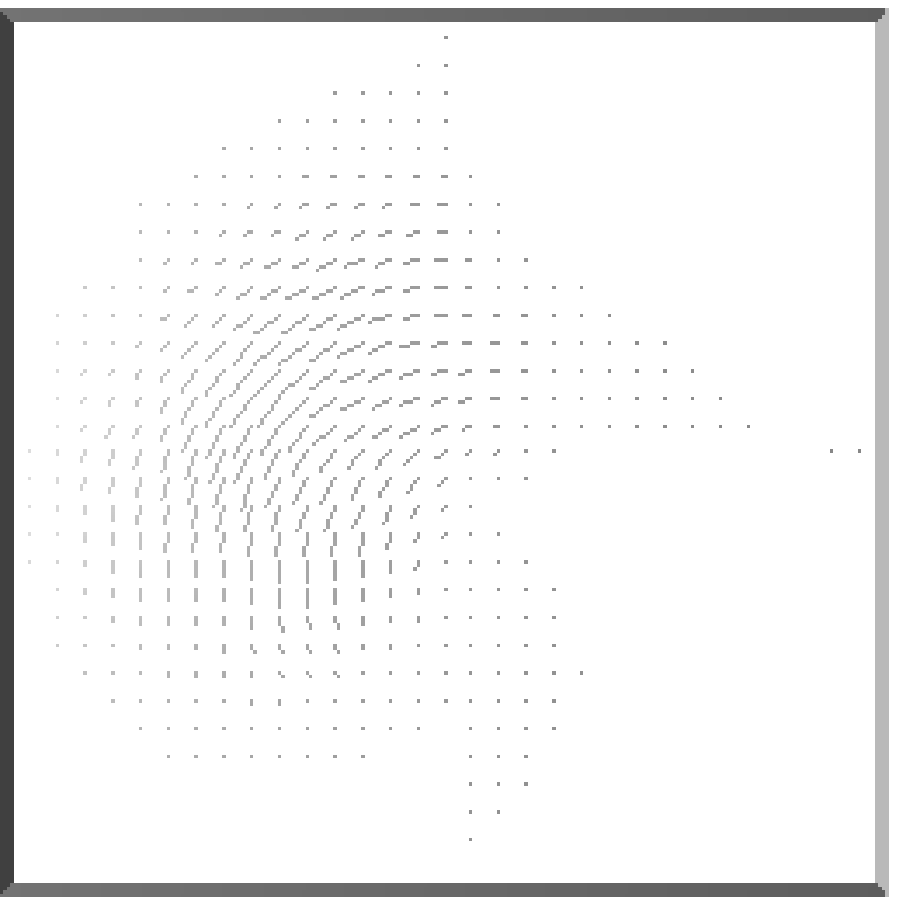}} &
      \epsfxsize=20mm\mbox{\epsfbox{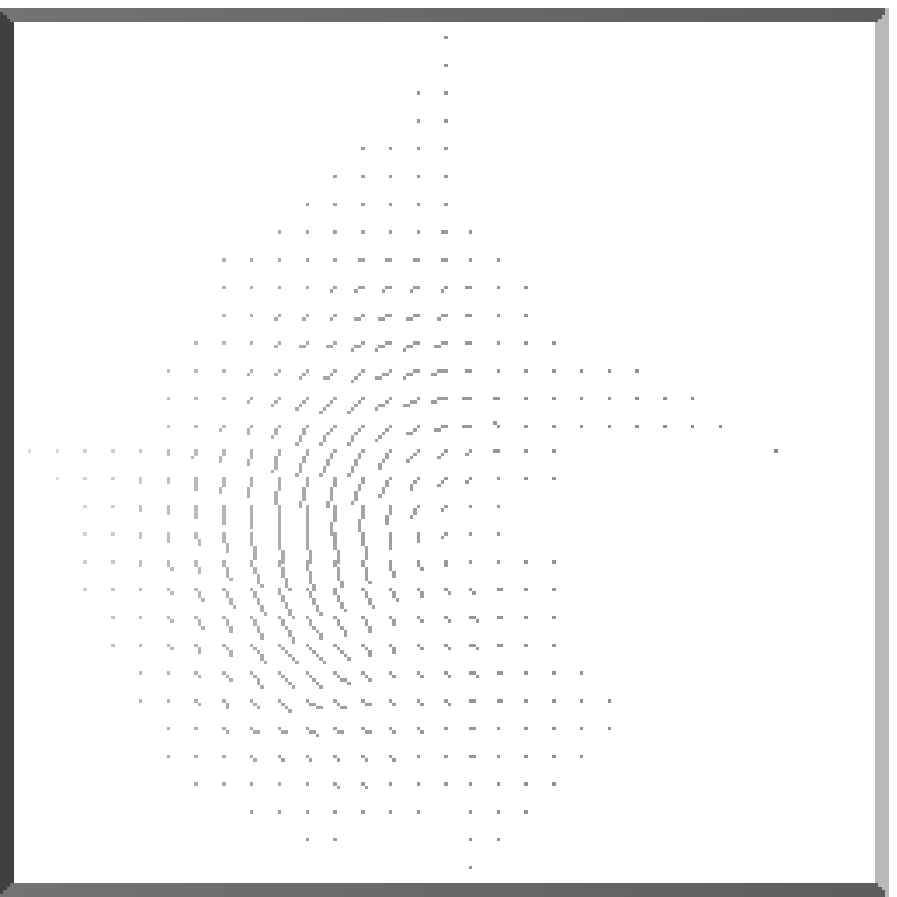}} &
      \epsfxsize=20mm\mbox{\epsfbox{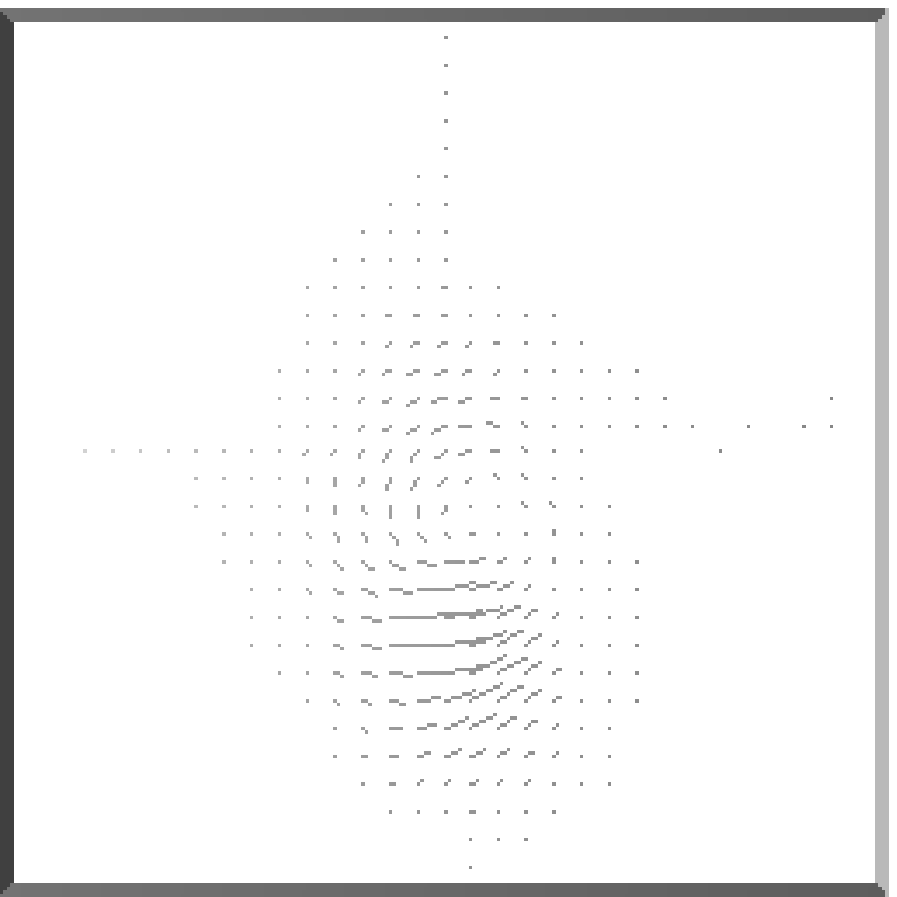}} &
      \epsfxsize=20mm\mbox{\epsfbox{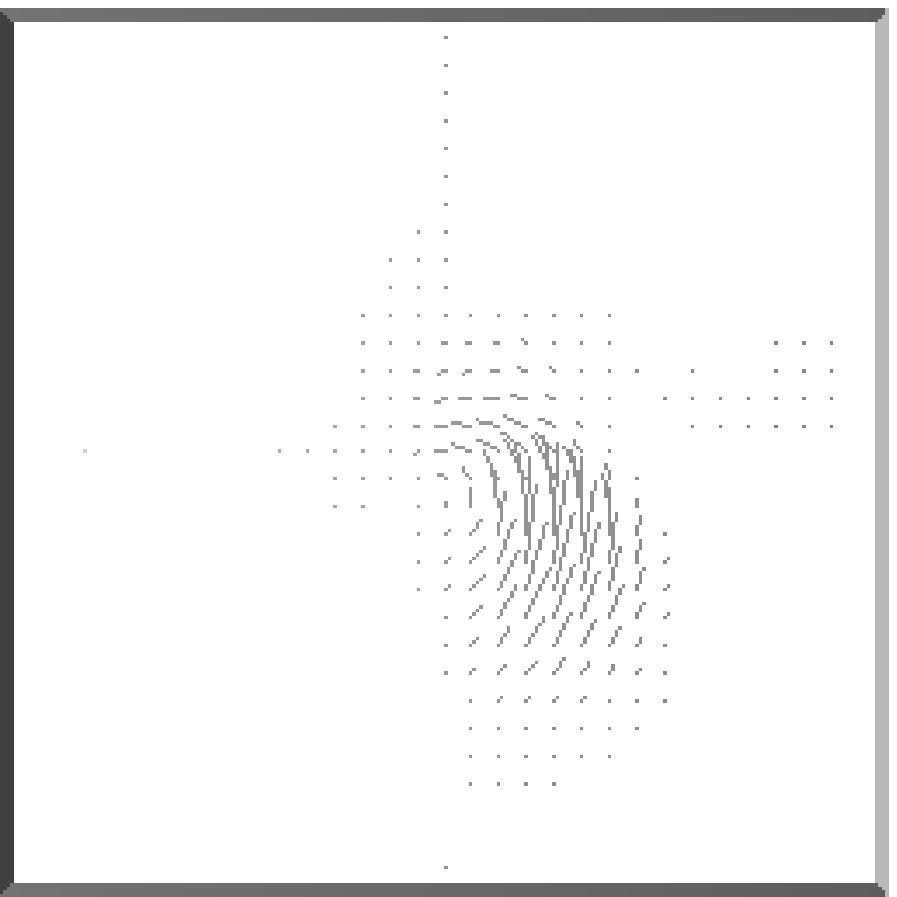}} \\
      t=12/8 & t=15/8 & t=18/8 & t=21/8 \\\\
    \end{tabular}
  \end{center}
  \caption{
    The evolution of the density and the current vector.
    The Gaussian is observed to circle around.
  }
  \label{fig3-2-2}
\end{figure}

\begin{figure}[H]
  \begin{center}
   \epsfxsize=50mm\mbox{\epsfbox{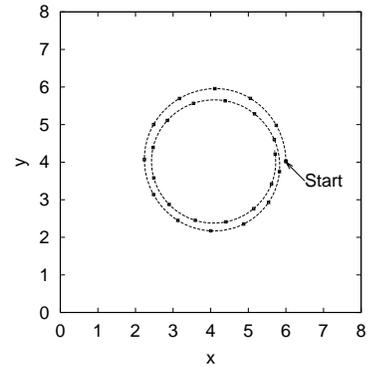}}
  \end{center}
  \caption{%
    The orbit of the average location of the wave packet.
    The radius of this circular trace is estimated as $2.0\text{a.u.}$
    The initial average location and momentum of this Gaussian
    are set as $(6\text{a.u.},4\text{a.u.})$
    and $(0\text{a.u.}, 4\text{a.u.})$, respectively.
  }
  \label{trace1}
\end{figure}

Some snapshots of the simulation time span are illustrated
in Fig.~\ref{fig3-2-2}.
The average location of the wave packet is observed to circle around
as plotted in Fig.~\ref{trace1}.


This trace is not a perfect circle but a swirl
due to the reflection by the closed walls around the system.

\begin{figure}[H]
  \begin{center}
    \epsfxsize=50mm\mbox{\epsfbox{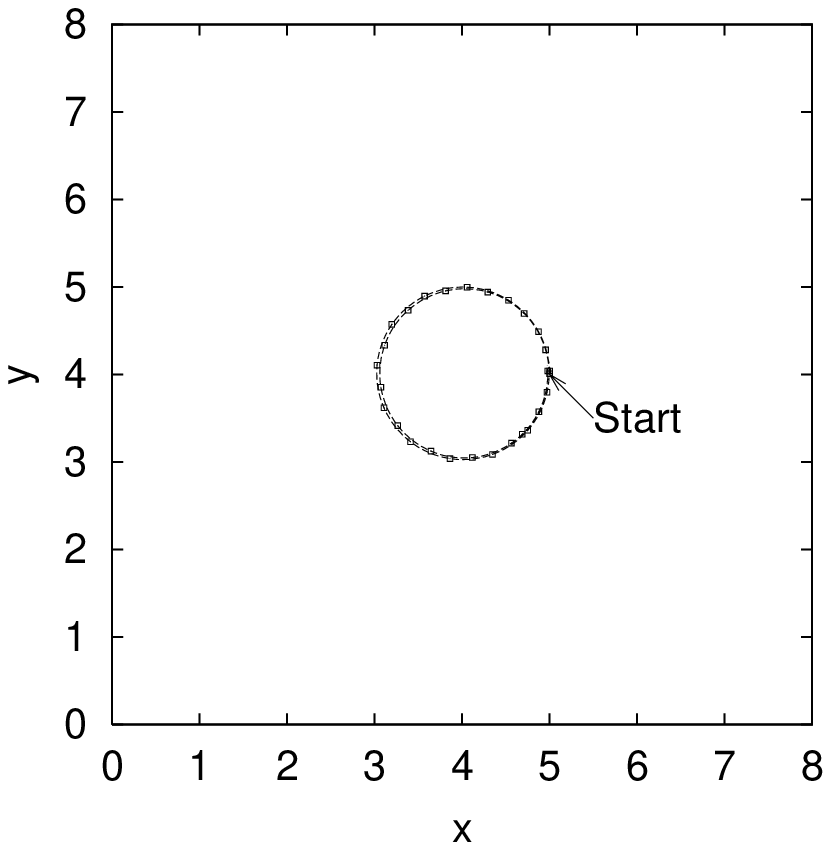}}
  \end{center}
  \caption{%
    The another orbit of the average location of the wave packet.
    The radius of this circular trace is estimated as $1.0\text{a.u.}$
    The initial average location and momentum of this Gaussian
    are set as $(5\text{a.u.},4\text{a.u.})$
    and $(0\text{a.u.}, 2\text{a.u.})$, respectively.
  }
  \label{trace2}
\end{figure}

A more perfect circular trace is observed by enlarging the system
or shortening the cyclotron radius to reduce the effect of the reflection.
Figure~\ref{trace2} shows the result of another simulation.
These results afford good agreement with the result by
classical mechanics.

\subsection{Aharonov-Bohm effect}

\begin{figure}[H]
  \begin{center}
    \epsfxsize=50mm\mbox{\epsfbox{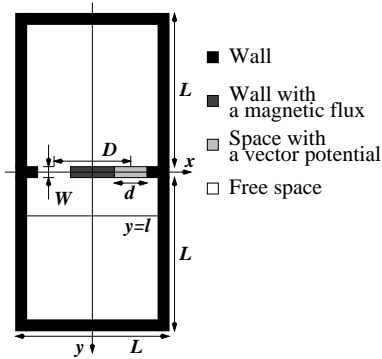}}
  \end{center}
  \caption{%
    The model system for the Aharonov-Bohm effect.
    The shape of this system is rectangular.
    A double-slit lies at the center.
    A magnetic flux $\Phi$ goes through a wall lying between the slits.
    $64\times 128$ computational grid points are allocated in the physical size
    $8\text{a.u.}\times 16\text{a.u.}$
    The initial wavefunction is set
    as a plane wave $k$ in front of the double-slit.
    The time slice is set as $\Delta{t}=1/64\text{a.u.}$
  }
  \label{fig3-3-1}
\end{figure}

We demonstrate Aharonov-Bohm effect by
simulating an electron dynamics on a system as illustrated
in Fig.~\ref{fig3-3-1}.

The vector potential is constructed as follows:
\begin{equation}
  {\bf A}(x,y)
  =
  ( 0,\ A_y(x),\ 0)^{\rm T}\ ;\,
  A_y(x)
  =
  -\int_{-L/2}^{x}\!\!\!\!\!\!\!\!{\rm d}x^\prime\,B(x^\prime,y)\ .
  \label{3-3-1}
\end{equation}
Thus $A_y(x)$ has a finite value only inside the right slit:
\begin{equation}
  A_y(x)
  =
  \left\{
    \begin{array}{ccl}
      -B(D-d) & : & \mbox{\rm inside the upper slit.} \\
       0      & : & \mbox{\rm in other area.}
    \end{array}
  \right.\ ,
  \label{3-3-2}
\end{equation}
where $d$ and $D$ mean the width of the slits and the span of the slits
respectively. Thus $D-d$ is the length of the wall
where a magnetic flux goes through.

In an analogy to semi-classical photon interference,
the electron interference pattern $I(x)$ in this AB system is
approximately described by the following form:
\begin{equation}
  I(x)
  \propto
  \Biggl|
    \frac{2\ell}{kdx}
    \sin{\Bigl[\frac{kd}{2\ell}x\Bigr]}
    \cos{\Bigl[\frac{kD}{2\ell} x
      - \frac{e\Phi}{2\hbar} 
    \Bigr]}
  \Biggr|^2\ .
  \label{3-3-3}
\end{equation}
In the above, $\ell$ is a $y$ coordinate where the pattern is evaluated.

\begin{figure}[H]
  \begin{center}
    \epsfxsize=50mm\mbox{\epsfbox{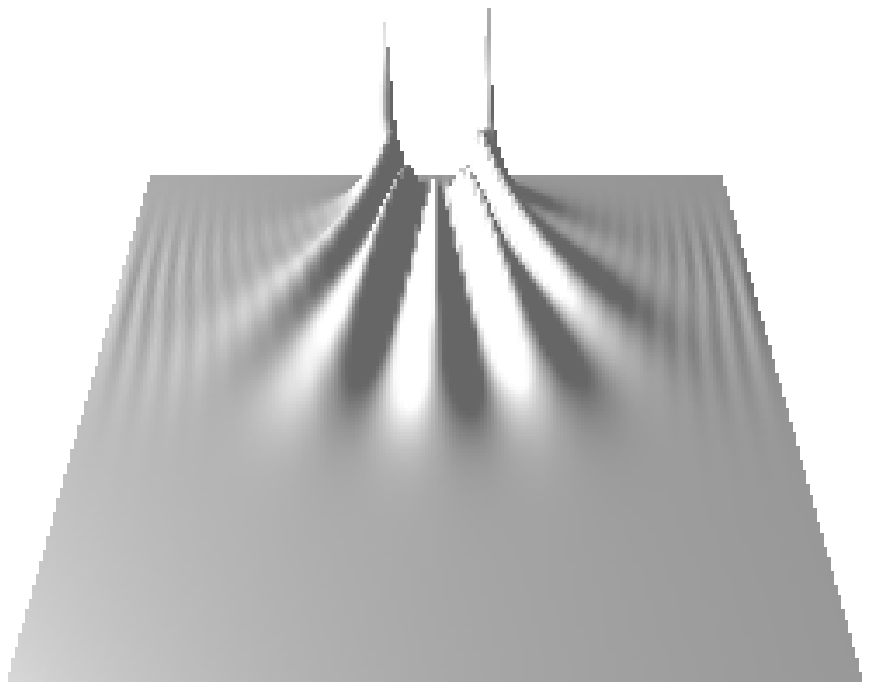}}\qquad
    \epsfxsize=60mm\mbox{\epsfbox{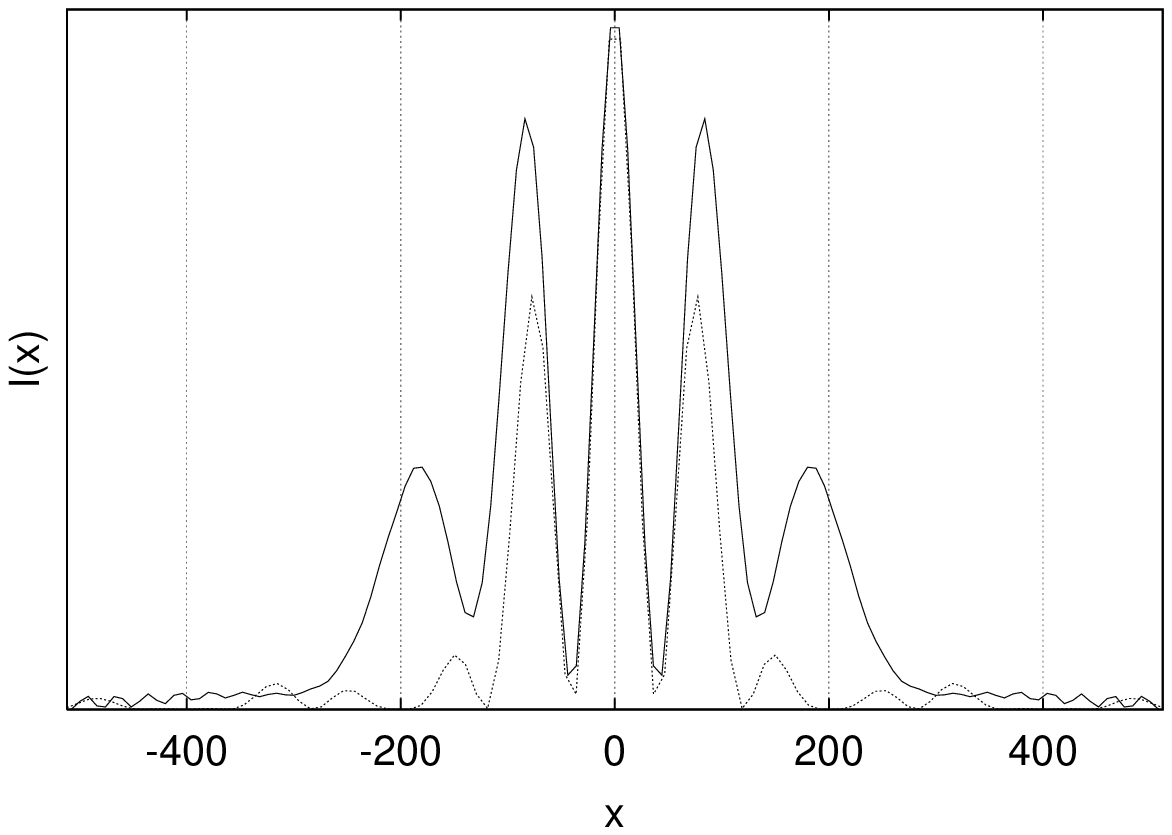}}
  \end{center}
  \caption{%
    The interference pattern observed in the back of the double-slit and
    at the line $y=\ell=L/4$ in a case of no magnetic flux, $\Phi=0$.
    The solid line indicates the numerical result; the dashed line indicates
    the semi-classical one derived from eq.~(\ref{3-3-3}).
  }
  \label{fig3-3-2}
\end{figure}

Figure~\ref{fig3-3-2} shows
the result of this simulation for the case of no magnetic flux, $\Phi=0$.
These data were taken soon after the pattern appeared
in order to prevent the pattern from extra interference
due to the reflected waves from side walls.
The interference pattern basically agrees with the semi-classical
one derived from eq.~(\ref{3-3-3}).

\begin{figure}
  \vspace*{-15mm}
  \begin{center}
    \epsfxsize=50mm\mbox{\epsfbox{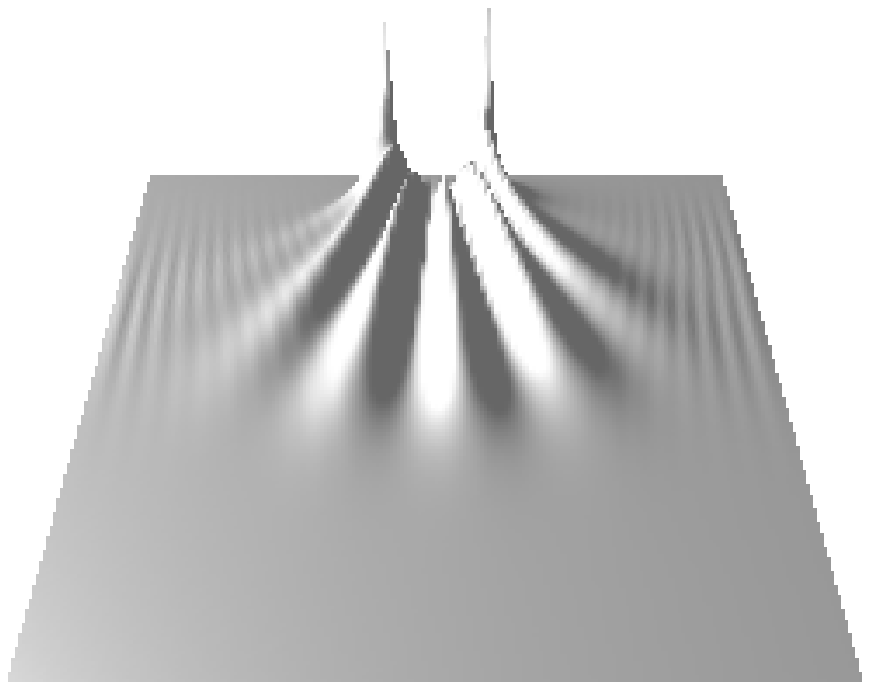}}\qquad
    \epsfxsize=60mm\mbox{\epsfbox{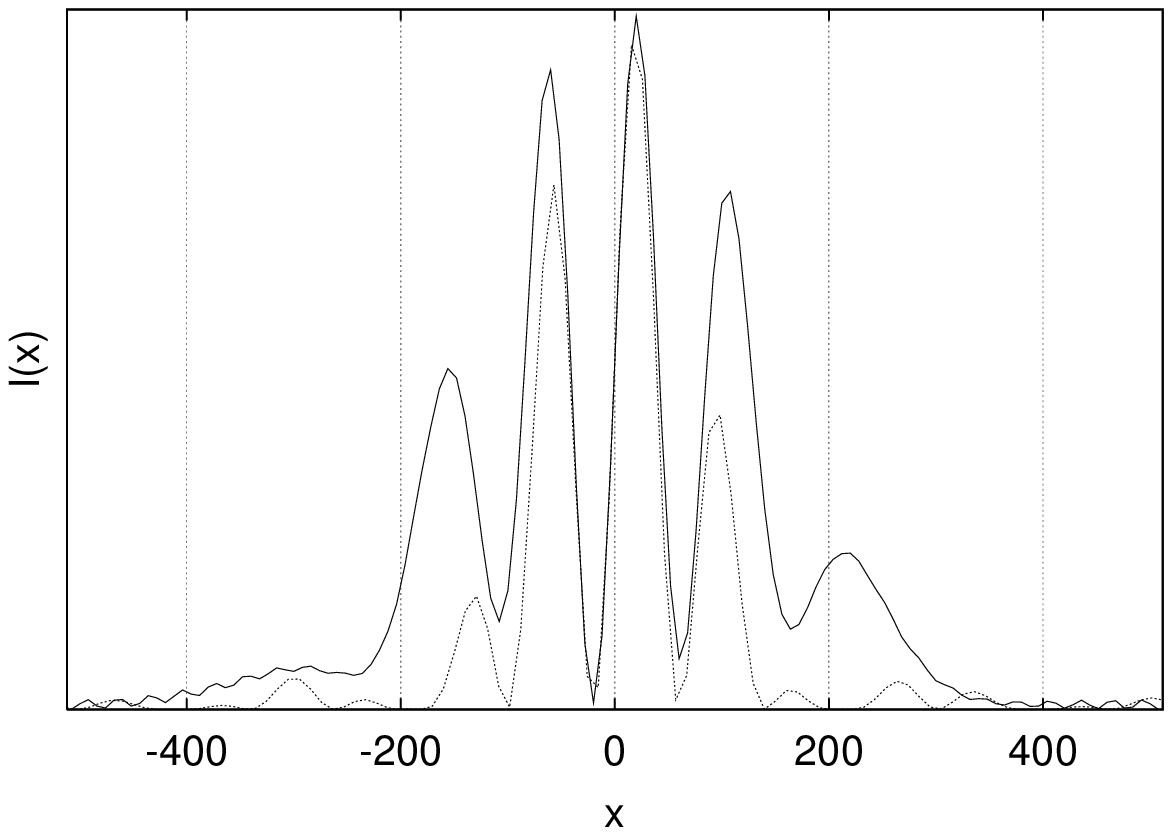}}
  \end{center}
  \caption{%
    The interference pattern observed in the back of the double-slit and
    at the line $y=\ell=L/4$ in a case of $\Phi=h/2e$.
    The solid line indicates the numerical result; the dashed line indicates
    the semi-classical one derived from eq.~(\ref{3-3-3}).
  }
  \label{fig3-3-3}
\end{figure}

\begin{figure}
  \vspace*{-15mm}
  \begin{center}
    \epsfxsize=50mm\mbox{\epsfbox{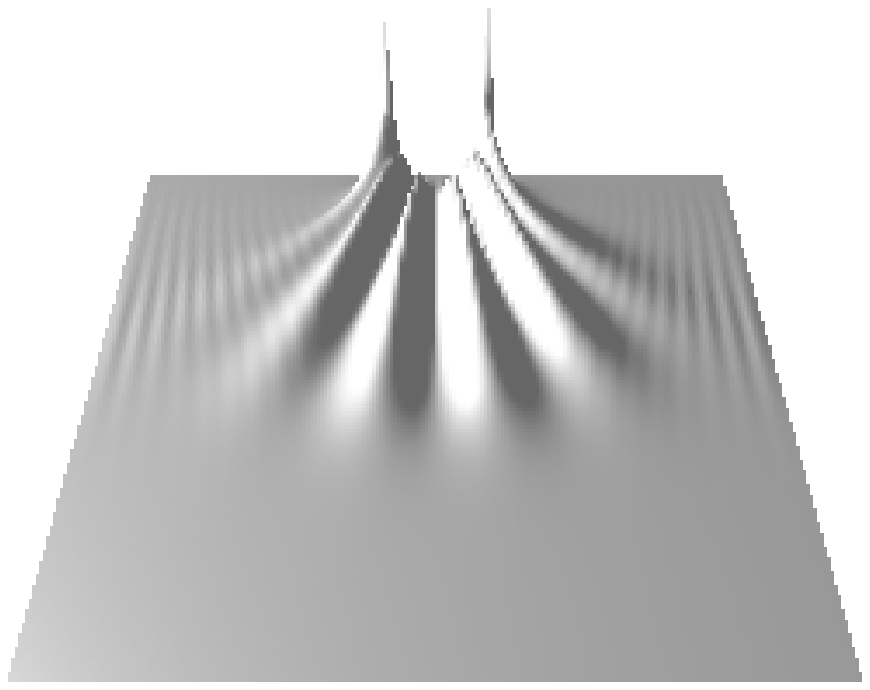}}\qquad
    \epsfxsize=60mm\mbox{\epsfbox{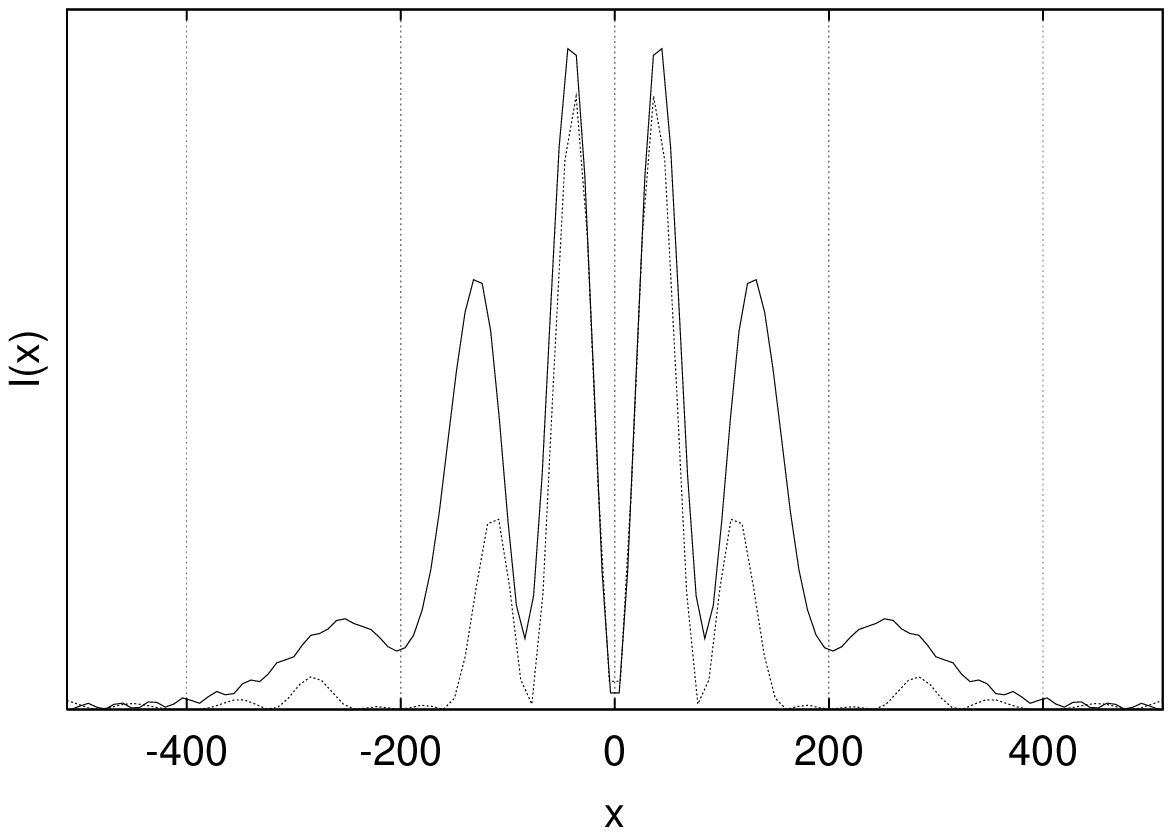}}
  \end{center}
  \caption{%
    The interference pattern observed in the back of the double-slit and
    at the line $y=\ell=L/4$ in a case of $\Phi=h/e$.
    The solid line indicates the numerical result; the dashed line indicates
    the semi-classical one derived from eq.~(\ref{3-3-3}).
  }
  \label{fig3-3-4}
\end{figure}

Further, the results for the case of magnetic flux $\Phi=h/2e$
and $\Phi=h/e$ are shown
in Figs.~\ref{fig3-3-3} and \ref{fig3-3-4}, respectively.
The patterns are observed to shift to the right-hand side, and these
behaviors also agree with the semi-classical one.
However, the patterns are different from the the semi-classical one
in their details. This is of course due to the quantum effect.


\section{Conclusion}

We have improved the computational method for the time-dependent
 Schr{\"o}dinger equation by utilizing the finite element method and
by formulating a new scheme for a magnetic field.
We have found that by using the FEM, the accuracy of the simulation
is dramatically improved without any increase in the computational cost.
We have also found that
the new scheme is quite efficient for simulating systems
in a magnetic field.

This computational method is especially useful for simulating
dynamics of electrons in a variety of meso-scopic structures.


\end{multicols}


\begin{thebibliography}{99}


\bibitem{Varga1962}
R. Varga, {\it Matrix Iterative Analysis} (Prentice-Hall, Englewood Cliffs, NJ, 1962), p.273.

\bibitem{DeRaedt1994}
H. De Raedt and K. Michielsen, Computers in Physics, {\bf 8}, 600 (1994).

\bibitem{Iitaka1994}
T. Iitaka: Phys. Rev. E {\bf 49} (1994) 4684.

\bibitem{Natori1997}
H. Natori and T Munehisa: J. Phys. Soc. Japan {\bf 66} (1997) 351.

\bibitem{Sugino1999}
O. Sugino and Y. Miyamoto: Phys. Rev. B {\bf 59} (1999)  2579.

\bibitem{Recipes}
W. H. Press, S. A. Teukolsky, W. T. Vetterling and B. P. Flannery:
{\sl Numerical Recipes in C} (Cambridge University Press, 1996)
chapter 19, section 2.

\bibitem{DeRaedt1994PRB}
H. De Raedt and K. Michielsen: Phys. Rev. B. {\bf 50} (1994) 631

\bibitem{Iitaka1997}
T. Iitaka, S. Nomura, H. Hirayama, X. Zhao, Y. Aoyagi and T. Sugano: Phys. Rev. E {\bf 56} (1997) 1222.

\bibitem{Kono1997}
H. Kono, A. Kita, Y. Ohtsuki and Y. Fujimura: J. Comput. Phys. (USA), {\bf 130} (1997) 148.

\bibitem{Tsuchida1998}
E. Tsuchida and M. Tsukada: J. Phys. Soc. Japan {\bf 67} (1998) 3844.

\bibitem{Watanabe2000}
N. Watanabe and M. Tsukada: Phys. Rev. E {\bf 62} No.2 (2000) {\sl in press}.

\bibitem{Suzuki1990}
M. Suzuki: Phys. Lett. A {\bf 146} (1990) 319. 

\bibitem{Suzuki1991}
M. Suzuki: J. Math. Phys. {\bf 32} (1991) 400.

\bibitem{Umeno1993}
K. Umeno and M. Suzuki: Phys. Lett. A {\bf 181} (1993) 387.

\bibitem{Suzuki1993}
M. Suzuki: Proc. Japan Acad. {\bf 69} Ser. B, 161 (1993). 

\bibitem{Suzuki1993springer}
M. Suzuki and K. Umeno: Springer Proceeding in Physics {\bf 76} (1993) 74.

\bibitem{Suzuki1995}
M. Suzuki: Phys. Lett. A {\bf 201} (1995) 425.

\end{thebibliography}
\end{document}